\documentclass[journal]{IEEEtran}
\usepackage{amsmath}
\usepackage{amsthm}
\usepackage{amssymb}
\usepackage{graphicx}
\usepackage{subcaption}
\usepackage{url}
\usepackage{epstopdf}
\usepackage{placeins}
\usepackage{epsfig}
\usepackage{amsfonts}
\usepackage{balance}
\usepackage{graphicx}
\usepackage{xfrac}
\usepackage{algorithm} 
\usepackage{algorithmic}
\usepackage{footnote}
\usepackage[table]{xcolor}

\newcommand{\algorithmicbreak}{\textbf{break}}
\newcommand{\BREAK}{\STATE \algorithmicbreak}
\makeatother

\newtheorem{theorem}{Theorem}

\newtheorem{example}{Example}
\newtheorem{remark}{Remark}

\newcommand{\mytilde}{\raise.17ex\hbox{$\scriptstyle\mathtt{‌​\sim}$}}
\usepackage{multirow}

\begin{document}
	\title{A Coded Caching Scheme with Linear Sub-packetization and its Application to Multi-Access Coded Caching}
	
		\author{%
			\IEEEauthorblockN{Anjana A. Mahesh, and B. Sundar Rajan\\}
			\IEEEauthorblockA{ Department of Electrical Communication Engineering, Indian Institute of Science, Bengaluru 560012, KA, India \\
				E-mail: \{anjanamahesh,bsrajan\}@iisc.ac.in}
		}

	{}

	\maketitle
	
	\begin{abstract} 
	This paper addresses the problem of exponentially increasing sub-packetization with the number of users in a centralized coded caching system by introducing a new coded caching scheme inspired by the symmetric neighboring consecutive side information index coding problem. The scheme has a placement policy where the number of sub-packets required grows only linearly with the number of users, with no restriction on file size, and a delivery policy which is instantaneously decodable. Further, an application of the new delivery scheme in a multi-access coded caching set-up is studied and a few results in that direction are presented. In particular, in the multi-access set-up, for cases where  optimality rate-memory trade-off characterizations are available, it is shown that the new delivery scheme  achieves optimal or near-optimal rates. 
	\end{abstract}


\IEEEpeerreviewmaketitle

\section{INTRODUCTION}
\label{sec:Intro}

\IEEEPARstart{I}{t} has been predicted in \cite{CISCO} that 82 percent of the global IT traffic will be video traffic by the year 2022. In \cite{CISCO}, the authors also report that peak hour traffic is growing at a much faster rate than average internet traffic and the main contributor to this accelerated peak hour traffic is video content which tends to have a "prime time". Caching has been proposed as a method to shift some of the peak hour traffic to off-peak hours by placing contents into caches across the network during off-peak hours. Since most of the video content is generated well ahead of transmission, this type of traffic sits well within the caching framework. 

In a seminal work \cite{MaN}, Maddah-Ali and Niesen introduced the notion of coding caching which jointly optimizes cache placement and content delivery by creating multi-cast opportunities and using coded transmissions to simultaneously deliver distinct contents to different users. The set-up in \cite{MaN}, which in this paper is called an $(N,K)$ centralized coded caching system, has a central server, which possesses $N$ files, coordinating cache placement as well as transmissions to $K$ different users, each possessing a dedicated cache memory which has a storage size of $M < N$ files. Since \cite{MaN}, researchers have explored several variants of the coded caching problem, including schemes with coded placement policies \cite{CFL,JV}, decentralized coded caching \cite{MaN_D}, device to device coded caching \cite{JCM}, online caching \cite{PMaN}, caching with non-uniform file popularity and demands \cite{NMa,ZLW}, multi-access coded caching \cite{HKD, ReK}, etc.

The placement phase of most of the coded caching schemes including \cite{MaN} involve splitting each file into $F$ sub-packets, where, $F$ increases exponentially with the number of users $K$. This exponentially increasing sub-packetization creates two problems: the number of bits needed to index the sub-packets and the need for large file sizes.  The first paper to look into the problem of sub-packetization was \cite{PDA} and since then, an array of papers \cite{PDA}-\cite{CKSB} have come up with coded caching schemes with reduced sub-packetization, a summary of which is given in \cite{CKSB}. Our paper also tries to address the sub-packetization problem by introducing a new coded caching scheme with linear sub-packetization inspired by the symmetric neighboring consecutive side information (SNCS) single unicast index coding problem introduced in \cite{MCJ}.  While the previous works which proposed schemes with sub-exponential or linear sub-packetization required the number of users to be either extremely large \cite{STD,CKSB} or take values in some specific restricted sets \cite{SZG, CKSB}, our scheme does not impose any restriction on the number of users or the file size.

Further, for an $(N,K,L)$-CCDN (Cache-aided Content Delivery Network) which, as defined in \cite{ReK}, is a multi-access coded caching system with a single server having access to $N$ files, $K$ users and $K$ caches, $N \geq K$ such that each user has access to $L$ consecutive caches, of storage size $M$ files, with a cyclic wrap-around, we show that our delivery scheme can be used as a transmission policy which will satisfy the demands of all users. The paper \cite{ReK} considers an $(N,K,L)$-CCDN at the memory points $M = \frac{iN}{K}, \ i \in \left[\left\lceil\frac{K}{L}\right\rceil\right] \cup \{0\}$ and for the $L \geq \frac{K}{2}$ regime, gives an upper and a lower bound on the achievable rate-memory trade-off (Corollary 2 and Theorem 3 in \cite{ReK} respectively) and exact optimality results for some special cases (Theorem 5).

The main technical contributions in this paper are listed below.

\begin{itemize}
	
\item For a centralized coded caching system with $K$ users and $N$ files, a cache-placement policy where the number of sub-packets grows linearly with $K$, irrespective of file size, is introduced.

\item For this placement scheme, a delivery policy which is instantaneously decodable (i.e, sub-packets involved in a transmission can be decoded by the users requesting them by utilizing only their cache contents and not any other transmission) is given. 

\item The rate-memory trade-off achieved by the delivery scheme is characterized.

\item We show that the delivery scheme introduced in this paper can be used in the delivery phase of a multi-access coded caching setting considered in \cite{ReK}, which was called an $(N,K,L)$-CCDN in \cite{ReK}. 

\item Based on the rate-memory trade off achieved by our delivery scheme, we give a new outer bound for the rate-memory trade-off of an $(N,K,L)$-CCDN with $L \geq \frac{K}{2}$. 

\item For the special cases at which the exact rate-memory was characterized in \cite{ReK}, we show that our delivery scheme either achieves the optimal rate or if doesn't, the gap from optimality goes to zero with increasing number of users.  
	
\end{itemize}

The rest of this paper is organized as follows. After stating the main results in this paper  in section \ref{sec:Results}, the new coded caching scheme is described in section \ref{sec:Scheme}. Then, after a brief review of the multi-access coded caching setting considered in \cite{ReK}, we describe how the delivery scheme in section \ref{sec:Scheme} can be used in the setting in \cite{ReK} and a few results in that direction in section \ref{sec:MAC}. Finally,  the paper is concluded in section \ref{sec:Conclusion}.

\emph{Notations}: For a prime power $q$, $\mathbb{F}_{q}$ denotes the finite field with $q$ elements. For a positive integer $n$, $[n]$ denotes the set $\{1,2,\cdots,n\}$. The set of positive integers is denoted by $\mathbb{Z}^+$. A $t$-subset of $[n]$ is a subset of $[n]$ of size $t$. The symbol $\oplus$ denotes the XOR of its operands. Also, $\binom{n}{k}= \frac{n!}{k!(n-k)!}$ and $\binom{n}{k}=0$, when $n < 1$ or $n < k$.  For an ordered set $\mathcal{S}$ with $n$ elements, the notation $\mathcal{S}(j)$ is used to denote the $j^{th}$ element in the set $\mathcal{S}$, for $j \in \{1,2,\cdots, n\}$. 

\section{Main Results}
\label{sec:Results}

\begin{theorem}
	For an $(N,K)$-coded caching system with cache memory size $M = \frac{iN}{K}$ for $i\in \{1,2,\cdots,K\}$, the rate 
	\begin{equation}
	R\left(\frac{iN}{K}\right) = 
	  \left\lceil\frac{K(K-i)}{2 + \left\lfloor\frac{i}{K-i+1}\right\rfloor + \left\lfloor\frac{i-1}{K-i+1}\right\rfloor}\right\rceil \cdot \frac{1}{K},
	\end{equation}
	
	is achievable with a linear sub-packetization $K$. 
	\begin{proof}
		This rate-memory pair can be achieved with a linear sub-packetization by using the placement policy described in subsection \ref{subsec:Placement} followed by the delivery scheme given in subsection \ref{subsec:Del2}. 
	\end{proof}

\end{theorem}

\begin{theorem}
	\label{Thm:MAC_UB}
Consider an $(N,K,L)$-CCDN with $L \geq \frac{K}{2}$. For cache memory size $M = \frac{N}{K}$, a new and improved upper bound for the rate $R(M)$ is given as 
\begin{equation}
\label{eq:MAC_UB_new}
R_{\text{UB}}(M) = \begin{cases}

K - \left( K - \left\lceil\frac{K(K-L)}{2 + \left\lfloor\frac{L}{K-L+1}\right\rfloor + \left\lfloor\frac{L-1}{K-L+1}\right\rfloor}\right\rceil \frac{1}{K}\right)\frac{MK}{N} 
\\   \hspace{4.2cm} \text{if } 0 \leq M \leq \frac{N}{K} \\

\left(\left\lceil\frac{K(K-L)}{2 + \left\lfloor\frac{L}{K-L+1}\right\rfloor + \left\lfloor\frac{L-1}{K-L+1}\right\rfloor}\right\rceil \frac{1}{K}\right) \left(2-\frac{MK}{N}\right)  \\ 
\hspace{4.2cm} \text{if } \frac{N}{K} \leq M \leq \frac{2N}{K} \\ 

 0  \hspace{4cm} \text{if } M \geq \frac{2N}{K}. 
\end{cases}
\end{equation}

\begin{proof}
	The $R_{\text{UB}}\left(\frac{N}{K}\right)$ is achieved by the delivery scheme in subsection \ref{subsec:Del2} as explained in section \ref{subsec:Del_MAC}. At $M=0$, worst-case rate is $K$ and for $L \geq \frac{K}{2}$, the rate at $M=\frac{2N}{K}$ is zero. The convex envelope of these three points is $R_{\text{UB}}(M)$.
\end{proof}
\end{theorem}
	
\section{Proposed Coded Caching Schemes}
\label{sec:Scheme}
In this section, we describe the proposed coded caching scheme with a linear sub-packetization. Consider an $(N,K)$ centralized coded caching system with $K$ users, $\mathcal{U} = \{1,2,\cdots, K\}$ and $N$ files, $\mathcal{W} = \{W_1, W_2, \cdots, W_N\}$. Each user has a dedicated cache which has a storage capacity of $M < N$ files.  The cache content at user $k$ is denoted as $\mathcal{Z}_k$ and the set $\mathcal{Z} = \left\{\mathcal{Z}_1, \mathcal{Z}_2, \cdots, \mathcal{Z}_K \right\}$ denotes the overall cache contents. The scheme is described for those cases where cache fraction, $\frac{M}{N}$, takes values in $\{\frac{1}{K}, \frac{2}{K}, \cdots, 1\}$. We now describe the placement phase followed by the delivery scheme.

\subsection{Placement Phase:}
\label{subsec:Placement}

Divide each file into $K$ sub-packets of equal size. The sub-packets corresponding to file $W_n$ are $\{W_{n,1}, W_{n,2}, \cdots, W_{n,K}\}$. Corresponding to a cache fraction of $\frac{M}{N} = \frac{i}{K}$, $i \in [K]$, the cache placement policy is as follows. The sub-packets $W_{n,k}, W_{n,k+1}, \cdots, W_{n,k+i-1}$ for all files $W_n \in \mathcal{W}$ are placed in the $k^{\text{th}}$ user's cache, i.e., for $k \in [K]$, $$\mathcal{Z}_k = \{W_{n,k}, W_{n,k+1}, \cdots, W_{n,k+i-1} | \ \forall \ W_n \in \mathcal{W}\}$$. 
Thus, each user has $i$ consecutive sub-packets, each of size $\frac{1}{K}$ units, of every file, in its cache. The memory occupied by these packets is $N \cdot \frac{i}{K}$ which is equal to $M$, thus satisfying the memory size constraint.

\subsection{Delivery Scheme}
\label{subsec:Del2}
 
Consider an $(N,K)$ centralized coded caching system with user cache memory M = $\frac{iN}{K}, \ i \in [K]$. Following the cache placement in subsection \ref{subsec:Placement} above, let the receiver $i$ demand the file $W_{d_i}$ and let $\mathbf{d} = \{d_1, d_2, \cdots, d_K\}$ denote the demand vector. Each receiver knows $i$ sub-packets of its demanded file and needs the remaining $(K-i)$ sub-packets. Thus, total number of demanded sub-packets is $K \cdot (K-i)$. For the demand vector $\mathbf{d}$, there exists a single unicast index coding problem \cite{BiK,BBJK,OHL} with number of users being equal to the number of demanded sub-packets where each user demands a distinct sub-file. Corresponding to each of the $K$ users in the coded caching set-up, there are $K-i$ receivers, each demanding a single sub-packet and all having the user's cache contents as side information, in the index coding problem. We give a delivery scheme which is an instantaneously decodable solution of the above index coding problem, i.e., we assume that the sub-packets involved in a particular transmission can be decoded, by the users requesting them, instantly upon receiving that transmission,  using their cache contents only and not any other transmission.  

Form the ordered set of demanded sub-packets $\mathcal{L} = \{W_{d_1,i}, W_{d_1,i+1}, \cdots, W_{d_1,K}, W_{d_2,i+1}, W_{d_2,i+2},\cdots, W_{d_2,1}, \\ \cdots, W_{d_k, i+k-1}, W_{d_k, i+k-2}, \cdots, W_{d_k, k-1}, \cdots, W_{d_K,i}, \\ W_{d_K, i+1}, \cdots, W_{d_K, K-1}\}$. An algorithm which takes $\mathcal{L}$, $K$ and $i$ as inputs and generate the codewords for transmission are given in Algorithm \ref{Algo:Scheme B} and the subroutines called in Algorithm \ref{Algo:Scheme B} are given in the Algorithms \ref{Algo:SubRoutine}, \ref{Algo:Update}, \ref{Algo:Rule} and \ref{Algo:Check}. In the notation $W_{d_{u+a},p+a}$ used in the algorithms, the addition in the user index $u$ as well as the addition in the sub-packet index $p$ is performed with wrap-around w.r.t $K$. An explanation of the procedure is given below.

\begin{algorithm}
	\caption{Algorithm for generating the transmissions}
	\begin{algorithmic}[1]
		\REQUIRE $\mathcal{L}$ and $K$, $i$.
		
		\STATE Define $\Gamma \triangleq K-i+1$.
		
		\STATE Define $t \triangleq 2 + \left\lfloor\frac{i}{\Gamma}\right\rfloor + \left\lfloor\frac{i-1}{\Gamma}\right\rfloor$.
		
		\STATE Generate the ordered set 
		\begin{align*}
		T_0 = \{\, & W_{d_1,i+1}, W_{d_2,1}, W_{d_{1+\Gamma},i+1+\Gamma}, W_{d_{2+\Gamma},1+\Gamma}, \cdots, \\ & \cdots, W_{d_{1+\left\lfloor\frac{i}{\Gamma}\right\rfloor\Gamma},i+1+\left\lfloor\frac{i}{\Gamma}\right\rfloor\Gamma}, W_{d_{2+\left\lfloor\frac{i-1}{\Gamma}\right\rfloor\Gamma},1+\left\lfloor\frac{i-1}{\Gamma}\right\rfloor\Gamma} \, \},
		\end{align*}
		with the $j^{\text{th}}$ element in $T_0$, i.e., $T_0(j)$ denoted as $W_{d_{u_j},p_j}$, where, $u_j$ is the user index and $p_j$ is the sub-packet index. 
		
		\IF{$t$ is odd AND ($i < K-2$)}
		
		\STATE $T_0(t) = W_{d_{u_t},p_t}$ is replaced by $W_{d_{u_t},p_t+1}$
		
		\ENDIF
		
		\STATE $\mathcal{C} = \emptyset$.
		
		\WHILE{$ |\mathcal{C}| \leq \left\lceil\frac{K(K-i)}{t}\right\rceil$}   
		
		\STATE	$T_1 = \emptyset$, flag = 0
		
		\FOR{each $W_{d_{u_j},p_j} \in T_0$}

		\IF{$W_{d_{u_j},p_j} \in \mathcal{L}$}
		
		\STATE	$T_1 = T_1 \cup W_{d_{u_j}, p_j}$.
		
		\ELSE
		
		\IF{$T_1 == \emptyset$ AND $|\mathcal{L}| == K$}
		\STATE $T_1$ = SubRoutine($\mathcal{L},K,i$)
		\ELSE		
		\STATE [$T_1$, flag] = Update($W_{d_{u_j},p_j}, \mathcal{L},K,T_1$, flag) 
		\ENDIF
		
		\ENDIF
		
		\ENDFOR
		
		\STATE $\mathcal{L} = \mathcal{L} \setminus T_1$
		
		\STATE $T_0 = \emptyset$.
		
		\STATE $c = 0$
		
		\FOR{each $W_{d_{u_j},p_j} \in T_1$}
		
		\STATE $c = c\bigoplus W_{d_{u_j},p_j}$
		
		\STATE $T_0 = T_0 \cup W_{d_{u_j+1},p_j+1}$
		
		\ENDFOR
		
		\ENDWHILE
		
		\RETURN Code $\mathcal{C}$ for transmission. 
	\end{algorithmic}
	\label{Algo:Scheme B}
\end{algorithm}

The first codeword is formed as 
\begin{equation}
\label{eq:Scheme B : Case 3}
c = \sum\limits_{\beta} \underbrace{W_{d_{1+\beta\Gamma},i+1+\beta\Gamma}}_{\text{term} 1} + \underbrace{W_{d_{2+\beta\Gamma},1+\beta\Gamma}}_{\text{term} 2}, 
\end{equation}
where, $\beta$ takes the values $0,1, \cdots \left\lfloor\frac{i}{\Gamma}\right\rfloor$ for term $1$ and the values ${0,1, \cdots \left\lfloor\frac{i-1}{\Gamma}\right\rfloor}$ for term $2$. Thus, the codeword $c$ is a sum of $t = 2 + \left\lfloor\frac{i}{\Gamma}\right\rfloor + \left\lfloor\frac{i-1}{\Gamma}\right\rfloor$ sub-packets. 

After generating a codeword, the corresponding sub-packets are removed from the set $\mathcal{L}$, and the next codeword is generated by incrementing both the file index and the sub-packet index by one for all the sub-packets involved in the current code-word. If a term not present in $\mathcal{L}$ appears, it is replaced by another sub-packet, the rules for choosing which are given in the Rule() subroutine in Algorithm \ref{Algo:Rule}. In some special cases, at the end of an iteration, when only $K$ sub-packets remain in $\mathcal{L}$, a special construction of the set $T_1$ is needed which is given by the subroutine in Algorithm \ref{Algo:SubRoutine}. This is especially the case when $i \leq \frac{K}{2}$ and $K-i$ is odd. 

Since, each codeword is the sum of $t$ elements and there are $K(K-i)$ sub-packets in $\mathcal{L}$, the number of transmissions required is $\Lambda = \left\lceil\frac{K(K-i)}{t}\right\rceil$. Hence, the codeword generation is continued for $\Lambda$ iterations. Each of the generated codeword is of size $\frac{1}{K}$ units and hence, the expression for the rate of transmission  $R(M)$ can be characterized as 

\begin{align}
\label{eq:Rate Scheme B}
R_{\text{New}}\left(\frac{iN}{K}\right) = \frac{\Lambda}{K}  = \left\lceil\frac{K(K-i)}{2 + \left\lfloor\frac{i}{K-i+1}\right\rfloor + \left\lfloor\frac{i-1}{K-i+1}\right\rfloor}\right\rceil \cdot \frac{1}{K}.
\end{align}

\begin{algorithm}
	\caption{SubRoutine($\mathcal{L}, K, i$)}
	\begin{algorithmic}[1]
		\REQUIRE $\mathcal{L}$ and $K$, $i$.
		
		\STATE $\Gamma = K-i+1$.
		
		\STATE $t = 2 + \left\lfloor\frac{i}{\Gamma}\right\rfloor + \left\lfloor\frac{i-1}{\Gamma}\right\rfloor$.
		
		\STATE $T = W_{d_1,k}$, for some $k \in [K]$ such that $W_{d_1,k} \in \mathcal{L}$. 
		
		\FOR{$j = 1$ to $t-1$}
		\STATE $T = T \cup W_{d_{1+\left\lfloor\frac{jK}{t}\right\rfloor},k+\left\lfloor\frac{jK}{t}\right\rfloor}$.\ENDFOR
		
		\RETURN $T$.
	\end{algorithmic}
	\label{Algo:SubRoutine}
\end{algorithm}

\begin{algorithm}
	\caption{Update($W_{d_{u_j},p_j}, \mathcal{L},\mathcal{Z},T_1$, flag)}
	\begin{algorithmic}[1]
		
		\IF{flag == 0}
		\FOR{k = 1:4}
		\STATE $\hat{x}$ = Rule($W_{d_{u_j},p_j}$,k)
		\IF{Check($\hat{x},\mathcal{Z},\mathcal{L},T_1$)}	
		\STATE $T_1 = T_1 \cup \hat{x}$
		\STATE flag = k
		\BREAK
		\ENDIF
		\ENDFOR
		\ELSIF{flag == 1 OR flag == 3}
		\STATE flag = flag + 1
		\STATE $T_1 = T_1 \cup$ Rule($W_{d_{u_j},p_j}$,flag) 
		\ELSIF{flag == 2 OR flag == 4}
		\STATE flag = flag - 1
		\STATE $T_1 = T_1 \cup$ Rule($W_{d_{u_j},p_j}$,flag)
		\ENDIF
		\RETURN $T_1$, flag
	\end{algorithmic}
	\label{Algo:Update}
\end{algorithm}

\begin{algorithm}
	\caption{Rule($W_{d_{u_j},p_j}$, flag)}
	\begin{algorithmic}[1]
		
		\IF{flag == 1}
		
		\STATE $\hat{x} = W_{d_{u_j},p_j+1}$.
		
		\ELSIF{flag == 2}
		
		\STATE $\hat{x} = W_{d_{u_j+1},p_j}$.
		
		\ELSIF{flag == 3}
		
		\STATE $\hat{x} = W_{d_{u_j-1},p_j}$.
		
		\ELSIF{flag == 4}
		
		\STATE $\hat{x} = W_{d_{u_j},p_j-1}$.
		
		\ENDIF
		\RETURN $\hat{x}$.
	\end{algorithmic}
	\label{Algo:Rule}
\end{algorithm}

\begin{algorithm}
	\caption{Check($W_{d_{u_j},p_j}, \mathcal{Z},\mathcal{L},T$)}
	\begin{algorithmic}[1]
		
		\IF{$W_{d_{u_j},p_j} \notin \mathcal{L}$}
		\RETURN $0$
		\ELSE
		\FOR{each $W_{d_{u_k},p_k} \in T$}
		\IF{$p_j \notin Z_{u_k}$ OR $p_k \notin Z_{u_j}$}
		\RETURN $0$
		\ENDIF
		\ENDFOR
		\ENDIF
		\RETURN $1$.
	\end{algorithmic}
	\label{Algo:Check}
\end{algorithm}

\begin{remark}
	\label{Rem:Case 1}
	When  $i = 1$ or $i = K-1$, for any $K$ and $N$, the placement is same as that in Maddah-Ali-Niesen scheme \cite{MaN} and  the delivery scheme in Algorithm \ref{Algo:Scheme B} gives the same set of transmissions as that in \cite{MaN}. Hence, the rate achieved at these points, i.e., $R\left(\frac{N}{K}\right) = \frac{K-1}{2}$ and $R\left(\frac{(K-1)N}{K}\right) = \frac{1}{K}$, are the same as that achieved by Maddah-Ali-Niesen scheme and hence optimal under the constraint of uncoded placement. 
\end{remark}

\begin{remark}
	\label{Rem:Case 2}
When $1 < i \leq \frac{K}{2}$ , for any $N$ and $K$, an instantaneously decodable transmission can only contain the combination of two sub-packets at the most.  The code given by Algorithm \ref{Algo:Scheme B} is given by 
\begin{equation}
\label{eq:del2_case2}
W_{d_{1+a}, i+a+k} + W_{d_{1+k+a},1+a},
\end{equation}
for $ \ a \in \{0,1,...K-1\}, \ k \in \{1,2,\left\lfloor\frac{K-i}{2}\right\rfloor\}$. 

The above equation gives $K\cdot\left\lfloor\frac{K-i}{2}\right\rfloor$ transmissions. 
When $(K-i)$ is odd, we also need to make the following set of transmissions.

\begin{equation}
\label{eq_del2_case2_odd}
W_{d_{1+a}, i+\left\lceil\frac{K-i}{2}\right\rceil+a} + W_{d_{\left\lfloor\frac{K}{2}\right\rfloor+1+a}, \left\lceil\frac{i}{2}\right\rceil+a},
\end{equation}
for $ a \in \left\{ 0,1,\cdots, \left\lceil\frac{K}{2}\right\rceil \right\}. $

The equation \eqref{eq_del2_case2_odd} above results in an additional $\left\lceil\frac{K}{2}\right\rceil$ transmissions when ($K-i$) is odd. 
Thus, for both $(K-i)$ odd as well as even, there is a  total  of $\left\lceil\frac{K(K-i)}{2}\right\rceil$ transmissions, each of size $\frac{1}{K}$, giving the rate $R(\frac{iN}{K}) = \left\lceil\frac{K(K-i)}{2}\right\rceil\cdot\frac{1}{K}$. 
\end{remark}

\textbf{Decodability} : Decodability of the above delivery scheme is explained by considering three cases as follows. \\
\textit{Case I : $i = 1$ or $i = K-1$} : For this case, as explained in Remark \ref{Rem:Case 1}, the transmissions given by Algorithm \ref{Algo:Scheme B} is a reordered form of the Maddah-Ali-Niesen scheme in \cite{MaN} and hence decodability is assured. \\

\textit{Case II :  $1 < i \leq \frac{K}{2}$} : For this case, consider the transmissions given by \eqref{eq:del2_case2}. In a single transmission, only two users, $u_1$ and $u_2$, with user indices $1+a$ and $1+k+a$ respectively, where $a$ and $k$ takes values as given in \eqref{eq:del2_case2} are involved. With respect to these users, the transmission in \eqref{eq:del2_case2} can be re-written as $W_{d_{u_1},u_2+i-1} + W_{d_{u_2},u_1}$. Since the user $u_j$ knows $i$  consecutive sub-packets with indices from $u_j$ to $u_j +i -1$, both the users involved in a transmission know the sub-packet not requested by it. Now, consider the transmission given in \eqref{eq_del2_case2_odd}. Here, $u_1 = 1+a$ and $u_2 = \left\lfloor\frac{K}{2}\right\rfloor+1+a$.  The user $u_1$ knows sub-packets with indices from $1+a$ to $i+a$ of all files, it knows $W_{d_{u_2},\left\lceil\frac{i}{2}\right\rceil +a}$. Similarly, $u_2$ knows the sub-packets $\left\lfloor\frac{K}{2}\right\rfloor+1+a$ to $\left\lfloor\frac{K}{2}\right\rfloor+i+a$ and hence knows $W_{d_{u_1},\left\lceil\frac{K-i}{2}\right\rceil+i+a}$. \\

\textit{Case III : $i > \frac{K}{2}$} : Consider the first codeword $c$ given in \eqref{eq:Scheme B : Case 3}. The terms involved in this codeword can be grouped into two sets as follows:

\begin{table}[H]
	\begin{center}
		\caption{Table containing terms in \eqref{eq:Scheme B : Case 3}.}
		\begin{tabular}{|c | c|}
			\hline
			Column 1 & Column 2 \\
			\hline\hline
			$ W_{d_1,i+1}$ & $W_{d_2,1}$ \\
		    $W_{d_{1+\Gamma},2}$  & $W_{d_{2+\Gamma},1+\Gamma}$ \\
			$W_{d_{1+2\Gamma},2+\Gamma}$ & $W_{d_{2+2\Gamma},1+2\Gamma}$ \\
			$W_{d_{1+3\Gamma},2+2\Gamma}$ & $W_{d_{2+3\Gamma},1+3\Gamma}$ \\
			$\vdots$  & $\vdots$ \\
			$W_{d_{1+ \left\lfloor\frac{i}{\Gamma}\right\rfloor \Gamma}, 2 + \left(\left\lfloor\frac{i}{\Gamma}\right\rfloor - 1\right)\Gamma}$ & $W_{d_{2+ \left\lfloor\frac{i-1}{\Gamma}\right\rfloor \Gamma}, 1 + \left\lfloor\frac{i-1}{\Gamma}\right\rfloor\Gamma}$\\
			\hline
		\end{tabular}
	\label{Tab:Scheme terms}
	\end{center}
\end{table}

A term in $c$ is of the form $W_{d_{u_j},p_j}$, where, $u_j$ is the user index and $p_j$ is the sub-packet index. If $\left\lfloor\frac{i}{\Gamma}\right\rfloor > \left\lfloor\frac{i-1}{\Gamma}\right\rfloor$, there are odd number of terms in $c$ and the last term is $W_{d_{1+ \left\lfloor\frac{i}{\Gamma}\right\rfloor \Gamma}, 3 + \left(\left\lfloor\frac{i}{\Gamma}\right\rfloor - 1\right)\Gamma}$ if $K-i > 2$ and $W_{d_{1+ \left\lfloor\frac{i}{\Gamma}\right\rfloor \Gamma}, 2 + \left(\left\lfloor\frac{i}{\Gamma}\right\rfloor - 1\right)\Gamma}$ if $K-i \leq 2$. If $\left\lfloor\frac{i}{\Gamma}\right\rfloor = \left\lfloor\frac{i-1}{\Gamma}\right\rfloor$, then there are even number of terms in $c$ and the last term is $W_{d_{2+ \left\lfloor\frac{i-1}{\Gamma}\right\rfloor \Gamma}, 1 + \left\lfloor\frac{i-1}{\Gamma}\right\rfloor\Gamma}$. In all these cases, the largest sub-packet index that appears in $c$ is $i+1$ corresponding to the first term $W_{d_1,i+1}$ and it can be verified that the largest user index that can appear in $c$ is also $i+1$. Similarly the smallest user-index as well as sub-packet index in $c$  is $1$. Hence $1 \leq u_j, p_j \leq i+1, \ j \in \{1,2,\cdots,t\}$. Because of this, it can be seen that the users in the first row, i.e, users $1$ and $2$, knows all the sub-packets except their own demanded sub-packet. Further, note that the user indices appear in $c$ in an increasing order $\{1,2,1+\Gamma, 2+\Gamma, \cdots, 1+\left\lceil\frac{i}{\Gamma}\right\rceil\}$. Similarly, the sub-packet indices except the first one, i.e., $i+1$, appears in an increasing order $\{1,2,1+\Gamma,\cdots, \}$

Now consider a general user in Column 1 of Table \ref{Tab:Scheme terms}, which is of the form $u_k = 1+k\Gamma$. This user knows all sub-packet indices from $u_k$ to $u_k+i-1$, i.e., from $1+k\Gamma$ to $1+k\Gamma+i-1$, which can be computed as $u_k+i-1 = 1+k\Gamma+i-1 =$\\
$k(K-i+1) + i = (k-1)\Gamma + K-i+1+i = (k-1)\Gamma + 1$, considering wrap-around w.r.t $K$. Since, the sub-packet indices appear in an increasing order, to show that user $u_k$ knows all the sub-packets in $c$, except its own demanded sub-packet, it suffices to show that it knows the sub-packets of its preceding and succeeding terms. It can be seen from Table \ref{Tab:Scheme terms} that the index of the sub-packet demanded by the user succeeding the user $u_k$, (which is the user with index $2+k\Gamma$), is $1+k\Gamma = u_k$ which is known to user $u_k$. Similarly, the preceding user index is $2+(k-1)\Gamma$ and its demanded sub-packet is $1+(k-1)\Gamma = u_k+i-1$. Hence, a general user in Column 1 can decode its required sub-packet from $c$ as it knows all the other sub-packets involved. 

Now, consider a general user $u_k$ in Column 2  of Table \ref{Tab:Scheme terms}, which is of the form $u_k = 2+k\Gamma$. This user knows sub-packets with indices $u_k$ to $u_k+i-1 = 2+(k-1)\Gamma$. The sub-packet demanded by the preceding and succeeding users of $u_k$ in $c$ are of indices $2+(k-1)\Gamma  =u_k +i-1$ and $2+k\Gamma = u_k$. Hence, similar to a general user in Column 1, a general user $u_k$ in Column 2 also knows all the sub-packets involved in $c$, except its own demanded sub-packet. Hence, we saw each of the $t$ users whose demanded sub-packets appear in a codeword $c$ knows all $(t-1)$ other sub-packets involved in $c$ and hence can decode their respective demanded sub-packet from the codeword using their cache contents. Since the other code-words are obtained by incrementing the user and sub-packet indices of all the terms in $c$ by equal amounts, the users involved in all the transmissions know all but its own demanded sub-packet.

When a term already removed from $\mathcal{L}$ in a previous iteration appears again, the Update() subroutine gives another sub-packet present in $\mathcal{L}$ by using one out of four rules. While doing this, the Check() subroutine is invoked to verify that the updated sub-packet is present in the caches of all the other users involved in that transmission and vice-versa. Hence, the code obtained from Algorithm \ref{Algo:Scheme B} is instantaneously decodable. 
%


\subsection{Illustrating Example}
To illustrate the placement and delivery schemes described above, we give the following example.

\begin{example}
\label{Example_CCScheme}
Consider an $(N=6, \ K=6)$-centralized coded caching system with $M =4$.
The cache contents and for the demand vector, $\mathbf{d}= \{1, 2,\cdots, 6\}$, the sub-packets requested by each user is shown in Table \ref{Tab:Ex_Scheme}.

\begin{table}[H]
	\begin{center}
		\caption{Example \ref{Example_CCScheme} - Cache contents and sub-packets required at each user for the demand vector $\mathbf{d} = \{1,2,3,4,5,6\}$.}
			\label{Tab:Ex_Scheme}
		\begin{tabular}{|c|c|c|}
			\hline
			User,$i$ & Cache contents, $\mathcal{Z}_i$ & Reqd. sub-packets \\[5pt]
			\hline \hline
			1 & $\{W_{n,1}, W_{n,2},W_{n,3},W_{n,4} \ | \forall n \in [6]\}$ & $\{W_{1,5}, W_{1,6}\}$ \\
			\hline
			2 & $\{W_{n,2},W_{n,3},W_{n,4}, W_{n,5}\ | \forall n \in [6]\}$ & $\{W_{2,6}, W_{2,1}\}$ \\
			\hline
			3 & $\{W_{n,3}, W_{n,4},W_{n,5},W_{n,6} \ | \forall n \in [6]\}$ & $\{W_{3,1}, W_{3,2}\}$ \\
			\hline
			4 & $\{W_{n,4}, W_{n,5},W_{n,6},W_{n,1} \ | \forall n \in [6]\}$ & $\{W_{4,2}, W_{4,3}\}$ \\
			\hline
			5 & $\{W_{n,5}, W_{n,6},W_{n,1},W_{n,2} \ | \forall n \in [6]\}$ & $\{W_{5,3}, W_{5,4}\}$ \\
			\hline
			6 & $\{W_{n,6}, W_{n,1},W_{n,2},W_{n,3} \ | \forall n \in [6]\}$ & $\{W_{6,4}, W_{6,5}\}$ \\
			\hline
		\end{tabular}
	\end{center}
\end{table}	
Delivery Scheme : The transmissions obtained from the scheme described in subsection \ref{subsec:Del2} are given below. 
\begin{align*}
& W_{5,4} + W_{1,5} + W_{2,1} + W_{4,2} \\
& W_{6,5} + W_{2,6} + W_{3,2} + W_{5,3} \\
& W_{1,6} + W_{3,1} + W_{4,3} + W_{6,2} 
\end{align*}
The rate achieved using this scheme is $R(2) = \frac{3}{6} = \frac{1}{2}$ with a sub-packetization of $K=6$. For the same set-up, Maddah-Ali-Niesen scheme \cite{MaN} attains a rate of $R_{\text{M-N}} = \frac{2}{5}$ with a sub-packetization of $\binom{6}{4} = 15$.

\end{example}


\section{Application of the Delivery Schemes for a Multi-Access Coded Caching System}
\label{sec:MAC}
 We now explain how the delivery scheme described in section \ref{sec:Scheme} can be employed as a transmission scheme for an $(N,K,L)$-CCDN and how the rate achieved by our scheme compares against the upper and lower bounds and optimality characterizations in \cite{ReK}.

\subsection{Delivery Scheme for an $(N,K,L)$-CCDN}
\label{subsec:Del_MAC}

For an $(N,K,L)$-CCDN, $L \in [K]$, with cache memory size $M = \frac{iN}{K}, i \in \{0,1,\cdots, \left\lceil \frac{K}{L} \right\rceil\}$, each user has access to $L$ consecutive caches and thus a total cache memory of $\frac{iLN}{K}$. This is similar to a centralized coded caching set-up with each user having a dedicated cache memory of $\frac{iLN}{K}$ except for the total amount of cache memory available in both the systems. In the $(N,K,L)$-CCDN, the total memory available is $K \cdot \frac{M}{N} = iN$, whereas, in the (N,K)-coded caching system, with $M= \frac{iLN}{K}$, it is $iLN$. 
Now, let us see how the delivery policy developed in section \ref{sec:Scheme} for an $(N,K)$ centralized coded caching system can be used as a delivery scheme for an $(N,K,L)$-CCDN.

\subsubsection{Case I - $L \geq \frac{K}{2}$} 
Let us first consider $L \geq \frac{K}{2}$. For this case, $\left\lceil \frac{K}{L} \right\rceil = 2$ and hence $M$ only takes values in $\left\{0, \frac{N}{K}, \frac{2N}{K}\right\}$ out of which the only non-trivial memory point is $M= \frac{N}{K}$ as at $M=0$, there is no cache memory and the worst-case rate is $K$, whereas  $R=0$  at $M = \frac{2N}{K}$. For $M= \frac{N}{K}$, the placement scheme in \cite{ReK} specializes as follows. Each file is split into $K$ equal parts and the $i^{\text{th}}$ sub-file is stored in the $i^{\text{th}}$ cache and user $k$ has access to the caches $k, k+1, \cdots, k+L$, which implies it has access to the sub-files indexed by  $k, k+1, \cdots, k+L$ of all files. This is exactly the same as the contents of the $k^{\text{th}}$ user's cache following the placement given in subsection \ref{subsec:Placement}. 

Now, for a  demand vector $\mathbf{d} = \{d_1, d_2, \cdots, d_K\}$, in both the settings, each user needs $K-L$ sub-packets of its demanded file. Thus, the index coding problem arising in both the settings are the same and hence admits the same solution. Thus, the delivery scheme in section \ref{sec:Scheme} is a valid solution for the multi-access coded caching setting in an $(N,K,L)$-CCDN with $M = \frac{N}{K}$ and $L \geq \frac{K}{2}$ as well.

\subsubsection{Case II - $L < \frac{K}{2}$} 
For this case, we need to consider the memory points $M = \frac{iN}{K}, i \in \{0,1,\cdots, \left\lceil \frac{K}{L} \right\rceil\}$ out of which worst-case rate $R=K$ at $M=0$ and $R=0$ at $M= \left\lceil \frac{K}{L} \right\rceil \cdot \frac{N}{K}$. At the remaining memory points, the placement scheme in \cite{ReK} splits each file into $F(i,L) = \binom{K-iL+i-1}{i-1}\cdot \frac{K}{i}$ subfiles. $F(i,L) = K$ only for $i = 1$ and $i = \left\lfloor \frac{K}{L} \right\rfloor < \frac{K}{L}$. For these two cases, placement in \cite{ReK} is same as our placement scheme as explained in Case I above. At all other values of $i$, the value of $F(i,L)$ is strictly greater than $K$. So, for a cache memory of $M =\frac{N}{K}$ as well as $M = \left\lfloor \frac{K}{L} \right\rfloor\frac{N}{K}$ in an $(N,K,L)$-CCDN with $L < K$, following the placement phase, corresponding to a demand vector, $\mathbf{d} = \{d_1, d_2, \cdots, d_K\}$, the index coding problem seen is the same as that in an $(N,K)$- coded caching network with cache memory $LM$ and hence our delivery scheme can be used for transmission. 

\subsection{New Improved Upper Bound for $L \geq \frac{K}{2}$}
\label{sec:New_UB}

\begin{figure}	
	\centering
	\includegraphics[trim = 20mm 0 0 0, scale=0.2]{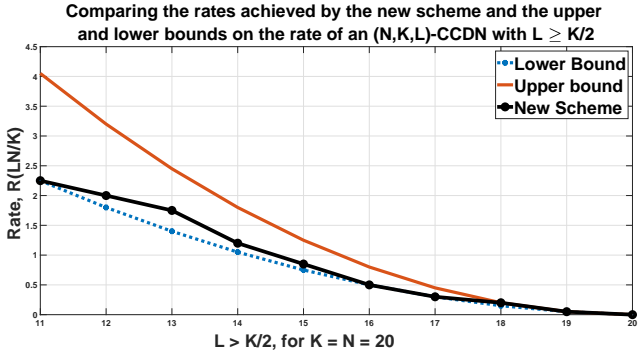}
	\caption{Rate-Memory trade-off Comparison}
	\label{fig:MAC_Compare}	
\end{figure}

A comparison plot showing the existing upper bound as well as lower bound and the rate achieved by the new scheme for the case $N=K=10$ and $L > \frac{K}{2}$ is given in Fig. \ref{fig:MAC_Compare}. Since our scheme performs better than the existing upper bound and since for $L < \frac{K}{2}$, the only bound on rate available is the upper bound (Theorem 1 in \cite{ReK}), the rate achieved by the newly introduced scheme in section \ref{sec:Scheme} can be used to derive an improved upper bound, which is given in Theorem \ref{Thm:MAC_UB}.  The improved upper bound based on our scheme is very close to the lower bound in general, and achieves it at some points. 

\subsection{Exact Optimality Cases}
\label{subsec:MAC_Optimality}

In \cite{ReK}, for exact optimality characterizations (Theorem 5), only three memory points, $M \in \left\{ 0, \frac{N}{K}, \frac{2N}{K}\right\}$ are considered.  Since the optimality results are given for the $L \geq \frac{K}{2}$ regime, among these three points, $M =\frac{N}{K}$ is the only non-trivial point as at $M=0$, there is no cache memory and the worst-case rate is $K$, whereas  $R=0$  at $M = \frac{2N}{K}$. A comparison between the optimal rate and that achieved by our scheme for the cases in Theorem 5 of \cite{ReK} is given in Table \ref{Tab:Optimality_Compare}. We can see that for all the cases covered,  our scheme either achieves the optimal rate-memory trade-off or the gap from optimality goes to zero with increasing number of users.

\begin{table}
	\begin{center}
		\caption{Comparison between the optimal rates (in Theorem 5, \cite{ReK}) and that achieved by the new scheme}.
			\label{Tab:Optimality_Compare}
		\begin{tabular}{|c|c|c|}
			\hline
			\multirow{2}{*}{$L$} & \multirow{2}{*}{$R^{*}\left(\frac{N}{K}\right)$} & \multirow{2}{*}{$R_{\text{New}}\left(\frac{N}{K}\right)$}\\
			& &  \\
			\hline\hline
			\multirow{2}{*}{$K-1$} & \multirow{2}{*}{$\frac{1}{K}$} & \multirow{2}{*}{$\frac{1}{K}$} \\
			& & \\
			\hline
			\multirow{3}{*}{$K-2$} & \multirow{3}{*}{$\frac{3}{K}$} & \multirow{3}{*}{$\begin{cases}
					\frac{3}{K} & \text{if } K = 3n, \ n \in \{2,3,\cdots,..\}\\
					\frac{4}{K} & \text{otherwise}
				\end{cases}$} \\
			& & \\
			& & \\
			\hline
			\multirow{6}{*}{K-3} & \multirow{6}{*}{$\frac{6}{K}$} & \multirow{6}{*}{$\begin{cases}
				\frac{9}{K} & \text{if } K = 6 \\
				\frac{8}{K} & \text{if } K = 5, 10 \\
				\frac{6}{K} & \text{if } K = 4n, \ n \in \mathbb{Z}^+ \\
				\frac{7}{K} & \text{otherwise}
				\end{cases}$} \\
			& & \\
			& & \\
			& & \\
			& & \\
			& & \\
			\hline
			\multirow{2}{*}{$K- \frac{K}{s} + 1$} & \multirow{3}{*}{$\frac{(K-s)}{2s^2}$} & \multirow{3}{*}{$\frac{(K-s)}{2s^2}$} \\
			&  & \\
			$s \in \mathbb{Z}^+$ & & \\
			\hline
			\end{tabular}
	\end{center}
\end{table}


\section{Concluding Remarks}
\label{sec:Conclusion}
In this paper, we looked at the problem of sub-packetization which poses a major challenge while implementing a coded caching protocol. Towards this end, we developed a coded caching scheme, with a sub-packetization that varies linearly with the number of users, inspired by the one-sided SNCS index coding problem. We also showed that the delivery scheme introduced in this paper is a solution for the index coding problem arising in the multi-access coded caching set-up called an $(N,K,L)$-CCDN  considered in \cite{ReK} and that the rate-expression corresponding to the newly introduced delivery scheme can be used to derive a new upper-bound for the rate achieved in an $(N,K,L)$-CCDN. Further, for the special cases for which the optimal rate-memory trade-off was characterized in \cite{ReK}, we analyzed the performance of our delivery scheme. 


\section*{Acknowledgment}
This work was supported partly by the Science and Engineering Research Board (SERB) of Department of Science and Technology (DST), Government of India, through J.C. Bose National Fellowship to Prof. B. Sundar Rajan.



\end{document}